\documentclass[epj]{svjour}

\usepackage[utf8]{inputenc}
\usepackage{graphicx}
\usepackage{amsmath}
\usepackage{amsfonts}
\usepackage[switch]{lineno}
\usepackage{color}
\usepackage[british]{babel}
\usepackage{cite}
\usepackage[absolute]{textpos}

\begin{document}

\title{Complex Langevin simulations and the QCD phase diagram: Recent developments}
\author{
Felipe Attanasio\inst{1} 
\and 
Benjamin Jäger\inst{2} 
\and 
Felix P.G.~Ziegler\inst{2}
}                     
\institute{
Department of Physics, University of Washington, Box 351560, Seattle, Washington 98195-1560, USA
\and 
CP3-Origins \& Danish IAS, Department of Mathematics and
Computer Science, University of Southern Denmark, Campusvej 55, 5230 Odense M,
Denmark}
\date{Received: date / Revised version: date}
\abstract{
In this review we present the current state-of-the-art on complex Langevin simulations and their implications for the QCD phase diagram. After a short summary of the complex Langevin method, we present and discuss recent developments. Here we focus on the explicit computation of boundary terms, which provide an observable that can be used to check one of the criteria of correctness explicitly. We also present the method of Dynamic Stabilization and elaborate on recent results for fully dynamical QCD.
\begin{textblock}{20}(10.4,0.3)
  CP3-Origins-2020-08 DNRF90\\
  NT@UW-20-27
 \end{textblock}%
} 
\maketitle

\section{Introduction}
\label{intro}

Strongly coupled quantum matter encompasses some of the most interesting problems in modern physics. Monte Carlo methods are commonly used to perform numerical simulations of theories not amenable to perturbative expansions. These methods typically rely on the path integral formulation of the theory in Euclidean space-time to have Boltzmann-like weights, which can be interpreted as probability distributions. This allows the generation of field configurations distributed according to these weights, whence observables can be sampled.

Real-time theories, QCD at finite baryon density, and non-relativistic bosons, amongst others, however, have complex actions and, therefore, complex weights in their path integrals. This forbids a probabilistic interpretation of the path integral measure. Moreover, this poses a big numerical challenge as oscillatory contributions from the generated configurations must cancel precisely in order to give accurate answers. This is known as the \textit{sign problem}.


In this work, we focus on the complex Langevin (CL) method. It is an extension of stochastic quantisation~\cite{Parisi:1980ys}, where real dynamical variables are allowed to take complex values. After some early works~\cite{Karsch:1985cb,Damgaard:1987rr}, it was realised that the method was plagued by runaway solutions and convergence to wrong limits~\cite{Ambjorn:1985iw,Klauder:1985ks,PhysRevB.34.1964,Ambjorn:1986fz}. More recently, complex Langevin simulations experienced a revival~\cite{Berges:2005yt,Berges:2006xc,Berges:2007nr,Pehlevan:2007eq,Aarts:2008rr,Aarts:2008wh,Guralnik:2009pk}, with studies focusing on understanding its properties and why it sometimes failed~\cite{Aarts:2009uq,Aarts:2010aq,Aarts:2010gr}. This progress led to the use of adaptive step size for the numerical integration~\cite{Aarts:2009dg}, which has improved the numerical stability and reduced the problem of runaway solutions. In addition, criteria for correctness~\cite{Aarts:2011ax,Nagata:2016vkn} that allow \textit{a posteriori} checks were formulated. Another byproduct of the resurgence of the complex Langevin method was the invention of the gauge cooling technique~\cite{Seiler:2012wz,Aarts:2013uxa}, inspired by gauge fixing~\cite{Berges:2007nr}, to limit excursions on the complex manifold in a gauge invariant way. Many more studies followed, applying the method to SU($3$) spin models~\cite{Aarts:2011zn}, the Thirring model~\cite{Pawlowski:2013pje,Pawlowski:2013gag}, random matrix theories~\cite{Mollgaard:2013qra,Mollgaard:2014mga,Bloch:2017sex}, and QCD with staggered quarks~\cite{Sexty:2013ica}, with a hopping expansion~\cite{Aarts:2014bwa}, and in the limit of heavy-dense quarks~\cite{Aarts:2016qrv}.
 
Additional uses of the complex Langevin method outside of QCD related models include superstring-inspired models~\cite{Nishimura:2019qal} and quantum many-body studies: rotating bosons~\cite{Hayata:2014kra,Berger:2018xwy}, spin-orbit coupled bosons~\cite{Attanasio:2019plf}, fermions with repulsive interactions~\cite{Loheac:2017yar}, non-zero polarisation~\cite{Loheac:2018yjh,Rammelmuller:2018hnk}, mass imbalance~\cite{Rammelmuller:2017vqn,Rammelmuller:2020vwc}, and to determine their virial coefficients~\cite{Shill:2018tan}. In addition, the phase structure of complex unitary matrix models \cite{Basu:2018dtm} as well as supersymmetric models and field theories \cite{Joseph:2019sof} has been studied.

\section{The Complex Langevin method}\label{sec:method}

\subsection{Overview}

The goal is to generate field configurations with a complex measure
\begin{equation}
e^{-S} \, \mathcal{D}\phi \equiv \rho\, \mathcal{D}\phi
\end{equation}
where $\phi$ generically represents all fields in the theory, and $S$ is a complex action defined on a real manifold $\mathcal{M}$.
This measure is replaced by a positive one, $P\, \mathcal{D}\phi_R \mathcal{D}\phi_I$, defined on the complexification $\mathcal{M}_c$ of $\mathcal{M}$.
This is the equilibrium measure of the Langevin process on $\mathcal{M}_c$.
When the criteria for convergence, outlined in~\cite{Aarts:2009uq,Aarts:2011ax}, are met observables calculated with either measure have the same expectation value.

The Langevin process is given by
\begin{align}
    d \phi_R &= K_R\, dt + dw\,,\\
    d \phi_I &= K_I\, dt\,,
\end{align}
where $t$ is known as the Langevin time, and $dw$ is a Wiener process normalised as $\langle dw^2 \rangle = 2 dt$. The drifts are given by
\begin{align}
    K_R &= - \text{Re} \left[ \nabla_\phi\, S[\phi_R + i \phi_I] \right] \,,\\
    K_I &= - \text{Im} \left[ \nabla_\phi\, S[\phi_R + i \phi_I] \right].
\end{align}
The choice to add the Wiener process only for the real field is arbitrary and can be generalised.
However, studies have shown that it is more beneficial to add noise only to the real part of the forces, see e.g.~\cite{Aarts:2013uza}. The process is said to produce correct results if the expectation value for a generic observable $\mathcal{O}$ for asymptotically long times, $\langle \mathcal{O} \rangle_\infty$, agrees with the `correct' expectation value, calculated with the original complex measure
\begin{equation}
    \langle \mathcal{O} \rangle_\infty = \langle \mathcal{O} \rangle_c \equiv \int \mathcal{D}\phi \,\, \mathcal{O}(\phi)\, e^{-S} \,. 
    \label{eq:long-time-identity}
\end{equation}
The complexification of the original manifold implies making all fields complex-valued.
The extra degrees of freedom can lead to unstable trajectories for the Langevin process.
Those can be stabilised, in some cases, by a change of variables, such as the non-trivial integration measure in the SU($3$) spin model, see ref.~\cite{Aarts:2012ft}.
In the case of gauge theories this means relaxing the unitary constraint of the gauge links, thus allowing the full space of non-singular matrices to be explored. For QCD this implies that the standard colour group SU($3$) is extended to SL$(3,\mathbb{C})$, which is not a compact group. Group elements can be arbitrarily far away from SU($3$), thus contradicting the criteria for correctness.
The method of gauge cooling~\cite{Seiler:2012wz,Aarts:2013uxa} was constructed as a means to bring the evolution closer to the unitary manifold in a gauge invariant way.

Gauge cooling consists of a series of gauge transformations constructed to reduce the distance to SU($3$) in a steepest descent fashion.
A recent discussion on how gauge cooling stabilises complex Langevin dynamics can be found in~\cite{Cai:2019vmt}.
In that work, the authors carry out analytical and numerical studies of the effects of gauge cooling on SU($2$) and SU($3$) Polyakov chains.
They point out four main effects of gauge cooling: 
(i) the removal of a large number of redundant degrees of freedom,
(ii) some components of the drift no longer have any effect on the dynamics,
(iii) emergence of
additional drift terms towards the unitary manifold supporting stability of the simulation, and (iv) the introduction of singularities in the drift.
(i) - (iii) stabilise the Langevin dynamics and support requirements for the criteria of correctness, while implications of (iv) remain still unknown. 
A justification for the gauge cooling method was provided in~\cite{Nagata:2015uga} for the continuum Langevin time formulation, whereas~\cite{Nagata:2016vkn} provides justification for discrete time.

\subsection{Criteria for Correctness} \label{sec.corr}
Correct convergence of the complex Langevin technique holds if the expectation value of an observable $\mathcal{O}$ over the probability distribution $P(\phi_R, \phi_I; t)$ agrees with that corresponding to the time-evolved complex measure $\rho(\phi; t)$ 
\begin{equation}
    \langle \mathcal{O} \rangle_{P(t)} = 
    \langle \mathcal{O} \rangle_{\rho(t)}\,.
\label{eq:CL-correctness}
\end{equation}
The important step in the  proof of correctness is the introduction of an interpolation function connecting the left and right-hand side of (\ref{eq:CL-correctness}) 
\begin{equation}
    F_{\mathcal{O}}(t,\tau) = \int \mathcal{D}\phi_R \mathcal{D}\phi_I \, P(\phi_R, \phi_I; t-\tau) \mathcal{O}(\phi_R, \phi_I; \tau)\,.
\label{eq:interpolf}
\end{equation}
Here the interpolating parameter $\tau$ is such that $0 \leq \tau \leq t$. For $\tau = 0$ (\ref{eq:interpolf}) reproduces the LHS of (\ref{eq:CL-correctness}) and for $\tau = t$ the RHS is recovered
assuming $P(\phi_R, \phi_I; 0) = \rho(\phi; 0)$, see equation (27) in~\cite{Aarts:2011ax} for a derivation. Correct convergence of the complex Langevin process requires the interpolating function
to be $\tau$-independent. This can be proven 
to hold in the 
absence of boundary terms arising from the integral (\ref{eq:interpolf}) (most prominently in $\phi_I$
-direction). This will be further discussed in the next section.

The formal argument of correctness relies on the holomorphicity of the action and hence of the drift. However, the justification of correctness
can be extended also to the presence of meromorphicity. For instance, in QCD zeros of the fermion determinant give rise to poles in the drift. 
Correctness can be ensured if the distribution $P$ 
vanishes sufficiently fast 
close to poles~\cite{Nishimura:2015pba,Aarts:2017vrv}.
In practice, this can be verified \textit{a posteriori}.
It should be noted, however, that a pole
lying inside the distribution $P$ can lead to a separation
of configuration space and to ergodicity problems.

On the level of the effective action 
complex Langevin studies of chiral random matrix theory~\cite{Mollgaard:2013qra,Splittorff:2014zca} 
and of effective Polyakov line models~\cite{Greensite:2014cxa} have identified the branch cut of the logarithm as a source of failure of correct convergence.
The ambiguity of the complex logarithm causes branch cut crossings, i.e.~a winding
of the CL trajectory around a pole in 
the drift.

This has been further investigated with regard to the criteria
of correctness in \cite{Nishimura:2015pba} by means of 
a model study on an action containing logarithmic 
singularities. 
Here, it was found that
the multi-valued character of the action is not 
the cause of the problem. 
The complex Langevin equation can be 
formulated in a single-valued fashion by deriving the 
drift from the weight $\rho(x)$ instead of the effective action.
The key to correctness, i.e.~the validity of
(\ref{eq:CL-correctness}) lies in the 
above mentioned behaviour of the 
probability distribution $P$ at 
and around the singularity 
corresponding to the branch point. 
As the authors show 
correct results can be obtained despite the occurrence of 
winding. Moreover, it is shown that in the case of a
single-valued action containing non-logarithmic singularities the complex Langevin 
method can lead to wrong results. In this case the 
probability distribution does not vanish sufficiently fast
close to the pole for (\ref{eq:CL-correctness}) to hold.

Complementary to that, extended analyses of simple models as well as simulations of full QCD 
provide further indications that the winding 
is not relevant for
the correctness of the method~\cite{Aarts:2017vrv}.

The criteria for correctness have to be checked for every observable. Care has to be taken in the presence of poles. Results are affected by the interplay 
of the pole order and the behaviour of the distribution $P$ and the observables around the pole. Recently, poles in the 
Complex Langevin equation have been studied in connection with boundary terms in \cite{Seiler:2020mkh}.

In \cite{Nagata:2016vkn, Nagata:2018net} the criteria of correctness are 
formulated from a slightly different but equivalent view point. There, the authors point out that the 
exponential (power-law) fall-off 
behaviour of the probability distribution associated with the drift indicates directly if CL works (fails). 
It is argued that this constitutes a necessary and sufficient criterion. The proof of the 
key identity (\ref{eq:long-time-identity}) is 
facilitated in terms of two ingredients. (i) formulating the correctness criteria 
(\ref{eq:CL-correctness})
in terms of a discretized Langevin time expansion
($\varepsilon$-expansion) of the time-evolved observable and the probability
distribution $P(\phi_R, \phi_I; t)$, and (ii) relaxing the assumption that the radius of convergence $\tau$ in the 
corresponding series expansion of the time evolution (\ref{eq:interpolf}) is infinite to being finite.
Correctness of the CL process follows if the 
$\varepsilon$-expansion is valid. The latter stands and falls with the exponential 
decay of the drift histogram. Moreover, it is interpreted that -- which is relevant for the next section -- 
the appearance of boundary terms arising from an integration by parts in the $\tau$-derivative
of the interpolation function is related to the breakdown of the $\varepsilon$-expansion.
The criterion is probed in numerical simulations 
using simple models \cite{Nagata:2016vkn}, 
gauge theories \cite{Nagata:2018net} and for full QCD \cite{Tsutsui:2019suq}.
For a comparison of the drift criterion within a study on correctness in terms of  
boundary terms see \cite{Scherzer:2018hid}.
Moreover, the validity of the complex Langevin method has
also been investigated recently from a 
mathematical perspective \cite{cai2020validity}.

\section{Recent developments}\label{sec:recent}

\subsection{Boundary terms}

The condition of (\ref{eq:interpolf}) being $\tau$-independent can be written as
\begin{equation}
    \frac{\partial}{\partial \tau} F_{\mathcal{O}}(t, \tau) = \lim_{Y\to\infty} B_{\mathcal{O}}(Y;t,\tau) = 0\,,
\end{equation}
where $Y$ is a cutoff in the non-compact directions and the middle term is a boundary term left from the integration
by parts
\begin{align}
    &B_{\mathcal{O}}(Y;t,\tau) =\nonumber\\
    &\int \mathcal{D}\phi_R \left[ K_I(\phi_R, Y) P(\phi_R, Y; t-\tau) \mathcal{O}(\phi_R + i Y; \tau) \right.\nonumber\\
    &\left. -K_I(\phi_R, -Y) P(\phi_R, -Y; t-\tau) \mathcal{O}(\phi_R - i Y; \tau) \right]\,.
    \label{eq:boundary-term}
\end{align}
This term has been recently investigated in~\cite{Scherzer:2018hid} for a simple one plaquette model with U($1$) symmetry.
In that model, (\ref{eq:boundary-term}) is non-zero and spoils the correctness of expectation values. Moreover it has been shown that the boundary term can be determined stochastically from the Langevin process.
Comparisons with numerical solutions of the Fokker-Planck equation (FPE), which describes the evolution of $P$, have been performed.
This is however very difficult in higher dimensions.
It is shown that the addition of a regulator term to the action is capable of reducing the boundary terms.

In~\cite{Scherzer:2019lrh}, studies of boundary terms have been deepened in the U($1$) one plaquette model and for the Polyakov chain. 
This has been also extended to field
theories such as the three-dimensional XY model as well as 
HDQCD.
In the case of gauge theories, the boundary terms appear to be related to the distance to the unitary manifold, typically measured via the unitarity norm~\cite{Seiler:2012wz}.
Moreover, it has been shown that boundary terms provide an estimate of the error of the Langevin process compared to the correct results. 
The quantification of the
latter relies on the assumption that the boundary term 
for the observables of interest is 
maximal at $\tau = 0$. From the analysis of the FPE this
was shown to hold in the $U(1)$ 
one-plaquette model for some classes of observables.
The error estimation demands the calculation of `higher order' boundary terms which are numerically more expensive.
The usefulness of this measure is currently seen to depend on the model being simulated and is under further investigation.
There are indications that the aforementioned assumption for the $\tau = 0$ behaviour is not valid for HDQCD.

\begin{table}
\centering
\caption{Boundary terms and the error of the complex Langevin simulations for the spatial plaquette in a HDQCD simulations with volume $6^4$, $\mu=0.85$, $N_F=1$, and $\kappa=0.12$, as shown in~\cite{Scherzer:2019lrh}.}\label{tab.BT}      
\begin{tabular}[!htb]{ccc}
\hline\noalign{\smallskip}
$\beta$ & $B$ & CL error \\
\noalign{\smallskip}\hline\noalign{\smallskip}
$5.1$ & $-0.578(22)$ & $0.056729(28)$  \\
$5.5$ & $-0.2808(99)$ & $0.020075(24)$  \\
$5.8$ & $-0.03058(14)$ & $-0.004869(54)$  \\
$6.0$ & $-0.00378(49)$ & $-0.000639(25)$  \\
\noalign{\smallskip}\hline
\end{tabular}
\end{table}
Table~\ref{tab.BT} summarises results from complex Langevin simulations of HDQCD from~\cite{Scherzer:2019lrh}.
It is worth noticing that the magnitude of $B$ is typically an order of magnitude larger than the error across a wide range of inverse couplings. The errors were computed by comparing the CL results with reweighting simulations. Table~\ref{tab.XY} shows results for the three-dimensional XY model, obtained using the dual worldline formulation~\cite{Banerjee:2010kc} and 
with the complex Langevin
method. The CL results were corrected using the boundary term analysis in~\cite{Scherzer:2019lrh}. The explicit computation of the boundary terms may hence serve to correct simulations with non-zero contribution from them.

In state-of-the-art full QCD simulations the boundary terms have to be monitored. In order to keep them small the CL trajectory is cut off when a prescribed threshold of the unitarity norm is reached\footnote{It has been recently shown in a study of 2D U($1$) gauge theory on a torus with a $\theta$-term that correct convergence can be obtained even when the unitarity norm is large~\cite{Hirasawa:2020bnl}. In that study, it was also found that observables only thermalise after the unitarity norm saturates.}~\cite{UBHD-68404032}. Moreover, decreasing $\beta$ can cause an increase in the boundary terms.

\begin{table}
\centering
\caption{Comparison of the worldline and the corrected complex Langevin simulations for the three-dimensional XY model, as shown in~\cite{Scherzer:2019lrh}.}\label{tab.XY}      
\begin{tabular}[!htb]{cccc}
\hline\noalign{\smallskip}
$\beta$ & $\mu^2$ & worldline & corrected CL \\
\noalign{\smallskip}\hline\noalign{\smallskip}
$0.2$ & $10^{-6}$ & $-0.062288(17)$ & $-0.06630(53)$ \\
$0.2$ & $0.1$ & $-0.062295(18) $ & $-0.06716(90)$ \\
$0.2$ & $0.2$ & $-0.062299(11) $ & $-0.0686(17)$ \\
\noalign{\smallskip}\hline\noalign{\smallskip}
$0.7$ & $10^{-6}$ & $-1.48219(35)$ & $-1.482283(34)$ \\
$0.7$ & $0.1$ & $-1.52398(35)$ & $-1.52399(72)$ \\
$0.7$ & $0.2$ & $-1.56641(20)$ & $-1.56476(48)$ \\
\noalign{\smallskip}\hline
\end{tabular}
\end{table}

\subsection{Dynamic stabilisation}

In ref.~\cite{Aarts:2016qrv} it was noticed that for some simulations the Langevin process would initially converge to the correct value (comparing with reweighting, when applicable) and then slowly drift to an incorrect one, despite the use of gauge cooling. The departure from the correct result always coincided with the increase in the distance of the field configurations from the unitary manifold. The method of dynamic stabilisation (DS) was then proposed. The idea is to add a term to the Langevin drift itself, which aims to  
(a) be small in comparison to the drift originating from the action; 
(b) vanish in the na\"ive continuum limit;
(c) affect only the non-compact directions of fields;
(d) be SU($3$) gauge invariant.

The additional force proposed in~\cite{Attanasio:2018rtq} is
\begin{equation}
    K_{x,\mu} \to K_{x,\mu} + i \alpha_{\mathrm{DS}} M_x,
\end{equation}
where $M_x$ is proportional to a power of the unitarity norm.
This choice of force, however, is non-holomorphic and thus violates the criteria of correctness.
The real parameter $\alpha_{\mathrm{DS}}$ controls the strength of the DS term. Given the complexity of gauge theories, it is difficult to predict its effects on the CL simulations \textit{a priori}, except in two limiting cases: when $\alpha_{\mathrm{DS}}$ is very small, the DS term will have a negligible effect and the dynamics should remain essentially unchanged; conversely, for large values of the control parameter the DS force heavily suppresses excursions into the non-unitary directions of SL($3,\mathbb{C}$), thus effectively re-unitarising the theory. The optimal values for $\alpha_{\mathrm{DS}}$ are found in a region where expectation values of the observables are least sensitive to it. Calculating the boundary terms explicitly could provide a non-heuristic way of determining optimal values of $\alpha_{\mathrm{DS}}$.

\begin{table}[]
\centering
\caption{Data comparing HMC vs. complex
Langevin (CL) simulations at vanishing chemical potential for two 
observables: plaquette and chiral condensate $\overline{\psi} \psi$.
These simulations used four flavours of na\"ive staggered fermions with
$\beta = 5.6$ and quark mass $a\,m_q= 0.025$ as shown in~\cite{Attanasio:2018rtq}.} \label{tab.DS}
\begin{tabular}{ccccc}
\hline\noalign{\smallskip}
				& \multicolumn{2}{c}{plaquette} & 
				\multicolumn{2}{c}{$\overline{\psi} \psi $} \\
				Volume & HMC & CL & HMC & CL \\
\noalign{\smallskip}\hline\noalign{\smallskip}
				$6^4$ & $0.58246(8)$ & $0.58245(1)$ & $0.1203(3)$ & $0.1204(2)$ \\
				$8^4$ & $0.58219(4)$ & $0.58220(1)$ & $0.1316(3)$ & $0.1319(2)$ \\
				$10^4$ & $0.58200(5)$ & $0.58201(4)$ & $0.1372(3)$ & $0.1370(6)$ \\
				$12^4$ & $0.58196(6)$ & $0.58195(2)$ & $0.1414(4)$ & $0.1409(3)$ \\
\end{tabular} 
\end{table}
Agreement between the complex 
Langevin method
and HMC at zero chemical potential can be found when Dynamic Stabilisation is used. Table~\ref{tab.DS} shows the accuracy that can be achieved. At vanishing chemical potential there is no sign problem, so that CL and HMC simulations should agree. However, the bi-linear noise scheme (see, e.g., \cite{Sexty:2013ica}) provides real drifts only on average. Thus a small source of complexity is always present, even though the theory itself is real. Without Dynamic Stabilisation this eventually leads to the failure of CL.

\subsection{Deformation technique}
When the quark chemical potential is sufficient for the formation of bound states, the fermion determinant exhibits zeros.
When the Langevin process explores regions around those zeros the drift becomes near-singular, leading to unstable simulations.
In ref.~\cite{Nagata:2018mkb} the fermion matrix is changed by the addition of a term $i\alpha \overline{\psi}(x)(\gamma_4 \otimes \gamma_4) \psi(x)$ to the Lagrangian density, where the $\gamma_4$'s act on spinor and flavour indices, respectively.
This deformation, for $\alpha$ large enough, moves the eigenvalue distribution of the fermion matrix away from zero.

Studies have been performed using a lattice of volume $4^3 \times 8$, coupling $\beta=5.7$, quark mass $am=0.05$, and chemical potential $0.4 \leq a\mu \leq 0.7$.
Histograms of the Langevin drift show a power-law behaviour for $\alpha < 0.3$, while the baryon number density and chiral condensate show a phase transition at $\alpha \sim 0.6$.
These observations indicate that an extrapolation from the deformed to the original manifolds should only use points simulated with $0.3 < \alpha < 0.6$.
Extrapolated values for the number density and chiral condensate simulated with CL were compared to RHMC simulations in the phase-quenched ensemble as a function of the chemical potential.
Both observables also show a steeper dependence on $\mu$ in the CL simulations, which is qualitatively consistent with the expectation that, in the thermodynamic limit at zero temperature, physical observables are independent of $\mu$ for $\mu < m_N/3$, for full QCD, or $\mu < m_\pi / 2$, in the phase-quenched case.

\subsection{QCD phase diagram}
 
Simulations of the QCD phase diagram with fully dynamical quarks are underway. Results have been reported for high~\cite{Sexty:2019vqx} and low~\cite{Tsutsui:2019suq,Ito:2018jpo,Kogut:2019qmi,Ito:2020mys} temperature regimes. The former two studies used four flavours of staggered quarks, while the latter used two flavours.
Additionally, a study of the deconfinement transition in QCD with heavy pions can be found in~\cite{Scherzer:2020kiu}, where two flavours of Wilson quarks have been considered.
All works have employed gauge cooling to reduce large explorations of non-unitary directions.

\subsubsection{High temperature}

\begin{figure}[]
\centering
\includegraphics[width=0.45\textwidth]{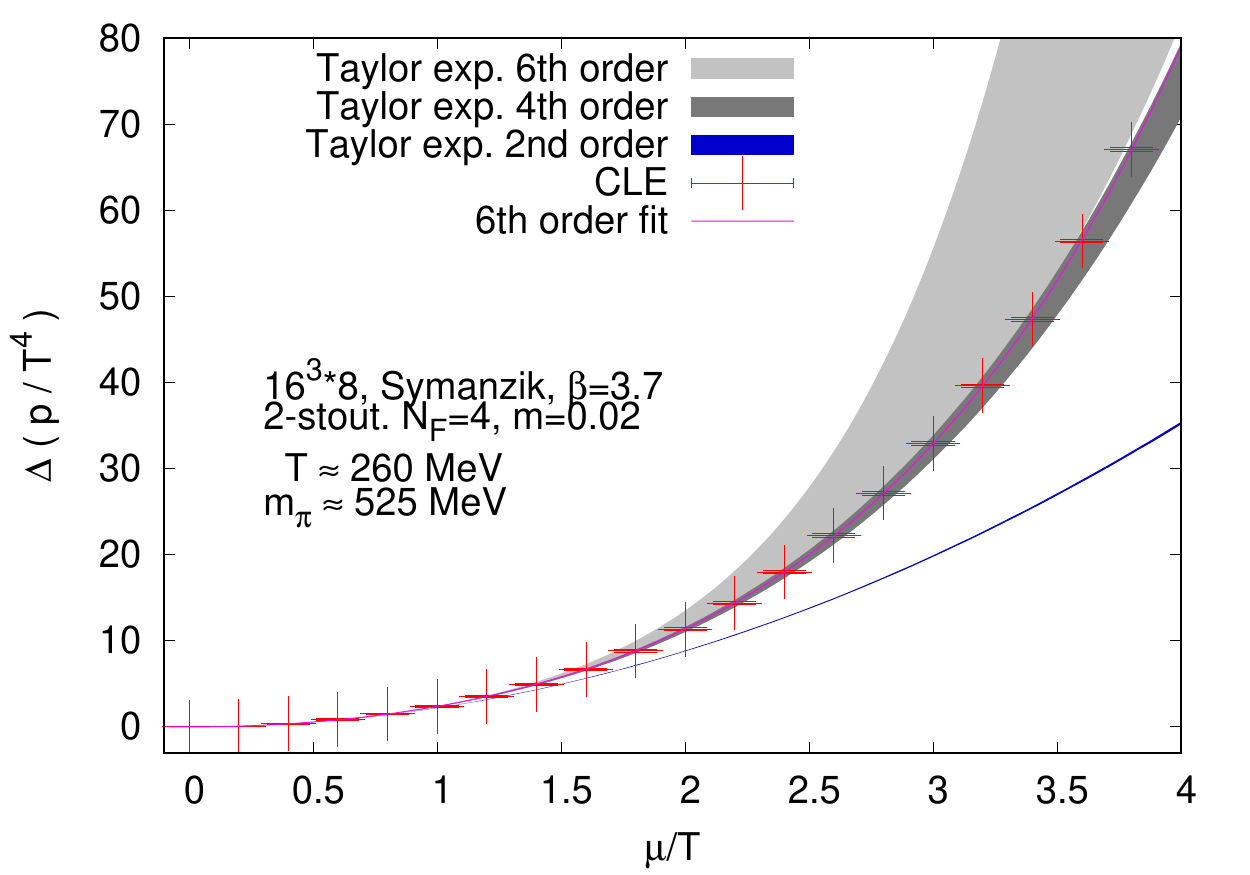}
\caption{\label{plot.Full} The pressure difference as a function of the chemical potential as shown in~\cite{Sexty:2019vqx}. The simulation parameters are listed in the figure.}\label{fig:Full}        
\end{figure}

In ref.~\cite{Sexty:2019vqx} the complex Langevin simulations were enhanced by stout smearing~\cite{Morningstar:2003gk} to smooth the gauge configurations. To achieve this, the smearing procedure had to be generalized to SL$(3,\mathbb{C})$ link variables. In addition, tree-level Symanzik improved gauge action was used for a volume of $16^3\times 8$. This allows a comparison between the standard Wilson plaquette action and the improved setup. The quarks are heavier than in nature to keep the simulations numerically feasible, resulting in a pion mass in the range of $500$ to $700\,$MeV. A comparison between standard Taylor expansion and complex Langevin simulation allows an additional consistency check. Figure~\ref{fig:Full} (left panel of figure 7 in~\cite{Sexty:2019vqx}) shows the pressure difference using a Taylor expansion up to 6th order as well as direct results from complex Langevin simulations. 
The agreement is remarkably good, so that complex Langevin simulations can be used to determine thermodynamic quantities, especially at high temperatures. 

\subsubsection{Low temperature}
In refs.~\cite{Ito:2018jpo,Tsutsui:2019suq,Ito:2020mys} the authors study a lower temperature setup with $\beta=5.7$ and $am=0.01$ for two lattice volumes: $8^3 \times 16$ and $16^3 \times 32$.
In order to ensure reliability of their results, 
they employed the criterion based on the distribution of the Langevin drift~\cite{Nagata:2016vkn}, described at the end of section \ref{sec.corr}.
After checking the gauge and fermion drifts in both lattices for difference values of the chemical potential, it has been determined that CL is expected to give correct results for $5.2 \leq \mu/T \leq 7.2$ for the smaller volume and $1.6 \leq \mu/T \leq 9.6$ for the larger one.

The observed plateau was understood qualitatively in terms of a picture of free fermions at zero temperature: due to the discreteness of the lattice momenta, $N_f N_c N_s$ is the maximum number of zero momentum quarks that can exist until the chemical potential is large enough to excite the first non-zero momentum states.
This interpretation is possible due to the smallness of the gauge coupling considered, making a picture of free fermions valid.
For a larger volume, they found that the plateau shifts to smaller values of $\mu$.
This is expected, as the discrete momentum states become closer.

In a complementary study of QCD at low temperature and finite chemical potential the authors of ref.~\cite{Kogut:2019qmi} used a setup of $\beta=5.6$ at volume $12^4$ and $\beta=5.7$ with volume $16^4$, both cases with quark mass $am=0.025$ and two flavours of staggered quarks for their CL simulations, augmented with gauge cooling.
They found that, in general, simulations at smaller coupling ($\beta=5.7$) produced results closer to the correct ones, when those were available, also reporting that the average unitarity norm decreases as the continuum limit is approached, in accordance with the findings of~\cite{Aarts:2016qrv}.
However, the expected transition from hadronic to nuclear matter at $\mu \approx m_N/3$, where $m_N$ is the nucleon mass, is not seen.
No signs of new, exotic phases of matter, such as colour-superconductors were found for $\mu \geq m_N/3$, and there is some indication of differences between the full theory and its phase-quenched approximation for $a\mu \geq 0.5$, but not large enough where saturation is dominant.

It is argued in ref.~\cite{Kogut:2019qmi} that the discrepancy between simulation results 
and physical expectations could be due to a number of issues complex Langevin simulations face.
One possibility is that CL is known to converge to phase-quenched results in some random matrix theories~\cite{Bloch:2017sex}, although this has not been observed in HDQCD.
Another is that CL produces correct results for non-singular observables, but those used in ref.~\cite{Kogut:2019qmi} do have poles.
This work uses two flavours of rooted staggered quarks. In~\cite{Kogut:2007mz} it has been argued that rooting is only applicable in a small vicinity of the continuum limit.

Lastly, for small temperatures and finite $\mu$, the CL faces severe problems because the eigenvalues of the Dirac matrix approach zero. Therefore the matrix becomes ill-conditioned and standard algorithms such as the conjugate gradient method break down. The latter represents the backbone of calculating the fermionic drift force.
In ref.~\cite{Bloch:2017jzi} the method of selected inversion has been suggested. It is a purely algebraic technique based on the sparse LU decomposition where a subset of the elements of the inverse matrix is computed, thus making it cheaper than a full inversion.

\subsubsection{Deconfinement transition}
In ref.~\cite{Scherzer:2020kiu} the phase diagram of QCD in the $T$-$\mu$ plane was investigated with two flavours of Wilson quarks, for chemical potentials up to $\mu \sim 5T$ using CL.
The study was carried out with a relatively high pion mass of $m_\pi \approx 1.3$ GeV, different lattice sizes, and focused on determining the deconfinement phase transition line.

The traditional parametrisation of the critical temperature for small chemical potentials by a polynomial was used.
By analysing the Binder cumulant of the Polyakov loop and of its fluctuations, it was possible to determine the curvature $\kappa_2$ of the transition line.
It has also been noticed that $\kappa_2$ has a non-monotonic behaviour as a function of the quark mass.
Despite the transition being a smooth crossover for the parameters considered
the critical temperature can be determined relatively well using the Binder cumulant.
A summary of the results is shown in Table \ref{tab:deconf}.

\begin{table}[]
\centering
\caption{
The fitted curvature and $T_c(0)$ for two flavour Wilson fermions with $N_s = 12$ and $16$ using two different methods taken from~\cite{Scherzer:2020kiu}. The curvature has the form $T_c(\mu) = T_c(0) - \kappa_2\, \frac{9\mu^2}{T_c(0)}$. More details on the two methods can be found in section IIIA and IIIB of~\cite{Scherzer:2020kiu}.} \label{tab.Deconf}
\begin{tabular}{cccc}
\hline\noalign{\smallskip}
Method & $N_s$ & $\kappa_2$ & $T_c(0)$/MeV\\
\noalign{\smallskip}\hline\noalign{\smallskip}
fit B3 &  $12$ & $0.001002(96)$ &  $303(2)$ \\
shift B3 & $12$ & $0.001167(55) $& $297(3)$ \\
\noalign{\smallskip}\hline\noalign{\smallskip}
fit B3 & $16$ & $8.1(2.4)\times 10^{-4}$&  $270(10)$ \\
shift B3 & $16$ & $0.001042(53)$ & $279(3)$\\
\end{tabular} 
\label{tab:deconf}
\end{table}

\section{Summary}

Recent developments on the complex Langevin method have shown promising progress. An important step has been to go beyond the analysis of the correctness criteria in simple models to a practical approach, which is applicable to field theories. With the determination of the boundary terms the error on the complex Langevin process can be estimated and, in some cases, compensated for. Challenges however remain in the development of this novel approach for full QCD. 

From a different angle, dynamic stabilization 
provides a viable technique for the CL to be applied to the 
regions of lower temperatures and medium densities.
Additionally, analysis of boundary terms, both numerically and analytically, may be able to provide a consistent way of optimising the control parameter of DS.

First results for CL simulations of the phase diagram of full QCD have recently appeared.
Despite the lack of evidence for the expected transition from hadronic to nuclear matter at zero temperature, results at low, but finite, temperate are encouraging and can be understood physically.
The agreement with Taylor expansion at high temperatures is a major step towards the ultimate goal of simulating the QCD phase diagram using the complex Langevin method.

\section{Acknowledgments}

We are grateful for discussion and collaboration with Gert Aarts and Ion-Olimpiu Stamatescu. We thank D\'enes Sexty for providing one of the figures to this manuscript.
The work of F.A. was supported by US DOE Grant No. DE-FG02-97ER41014 MOD27.

\bibliographystyle{epj}
\bibliography{main}

\begin{thebibliography}{70}

\bibitem{Parisi:1980ys}
G.~Parisi, Y.s. Wu, Sci. Sin. \textbf{24}, 483 (1981)

\bibitem{Karsch:1985cb}
F.~Karsch, H.W. Wyld, Phys. Rev. Lett. \textbf{55}, 2242 (1985)

\bibitem{Damgaard:1987rr}
P.H. Damgaard, H.~H{\"u}ffel, Phys. Rept. \textbf{152}, 227 (1987)

\bibitem{Ambjorn:1985iw}
J.~Ambjorn, S.K. Yang, Phys. Lett. \textbf{B165}, 140 (1985)

\bibitem{Klauder:1985ks}
J.R. Klauder, W.P. Petersen, J. Stat. Phys. \textbf{39}, 53 (1985)

\bibitem{PhysRevB.34.1964}
H.Q. Lin, J.E. Hirsch, Phys. Rev. B \textbf{34}, 1964 (1986)

\bibitem{Ambjorn:1986fz}
J.~Ambjorn, M.~Flensburg, C.~Peterson, Nucl. Phys. \textbf{B275}, 375 (1986)

\bibitem{Berges:2005yt}
J.~Berges, I.O. Stamatescu, Phys. Rev. Lett. \textbf{95}, 202003 (2005),
  \texttt{hep-lat/0508030}

\bibitem{Berges:2006xc}
J.~Berges, S.~Bors\'anyi, D.~Sexty, I.O. Stamatescu, Phys. Rev. \textbf{D75},
  045007 (2007), \texttt{hep-lat/0609058}

\bibitem{Berges:2007nr}
J.~Berges, D.~Sexty, Nucl. Phys. \textbf{B799}, 306 (2008), \texttt{0708.0779}

\bibitem{Pehlevan:2007eq}
C.~Pehlevan, G.~Guralnik, Nucl. Phys. \textbf{B811}, 519 (2009),
  \texttt{0710.3756}

\bibitem{Aarts:2008rr}
G.~Aarts, I.O. Stamatescu, JHEP \textbf{09}, 018 (2008), \texttt{0807.1597}

\bibitem{Aarts:2008wh}
G.~Aarts, Phys. Rev. Lett. \textbf{102}, 131601 (2009), \texttt{0810.2089}

\bibitem{Guralnik:2009pk}
G.~Guralnik, C.~Pehlevan, Nucl. Phys. \textbf{B822}, 349 (2009),
  \texttt{0902.1503}

\bibitem{Aarts:2009uq}
G.~Aarts, E.~Seiler, I.O. Stamatescu, Phys. Rev. \textbf{D81}, 054508 (2010),
  \texttt{0912.3360}

\bibitem{Aarts:2010aq}
G.~Aarts, F.A. James, JHEP \textbf{08}, 020 (2010), \texttt{1005.3468}

\bibitem{Aarts:2010gr}
G.~Aarts, K.~Splittorff, JHEP \textbf{08}, 017 (2010), \texttt{1006.0332}

\bibitem{Aarts:2009dg}
G.~Aarts, F.A. James, E.~Seiler, I.O. Stamatescu, Phys. Lett. \textbf{B687},
  154 (2010), \texttt{0912.0617}

\bibitem{Aarts:2011ax}
G.~Aarts, F.A. James, E.~Seiler, I.O. Stamatescu, Eur. Phys. J. \textbf{C71},
  1756 (2011), \texttt{1101.3270}

\bibitem{Nagata:2016vkn}
K.~Nagata, J.~Nishimura, S.~Shimasaki, Phys. Rev. \textbf{D94}, 114515 (2016),
  \texttt{1606.07627}

\bibitem{Seiler:2012wz}
E.~Seiler, D.~Sexty, I.O. Stamatescu, Phys. Lett. \textbf{B723}, 213 (2013),
  \texttt{1211.3709}

\bibitem{Aarts:2013uxa}
G.~Aarts, L.~Bongiovanni, E.~Seiler, D.~Sexty, I.O. Stamatescu, Eur. Phys. J.
  \textbf{A49}, 89 (2013), \texttt{1303.6425}

\bibitem{Aarts:2011zn}
G.~Aarts, F.A. James, JHEP \textbf{01}, 118 (2012), \texttt{1112.4655}

\bibitem{Pawlowski:2013pje}
J.M. Pawlowski, C.~Zielinski, Phys. Rev. \textbf{D87}, 094503 (2013),
  \texttt{1302.1622}

\bibitem{Pawlowski:2013gag}
J.M. Pawlowski, C.~Zielinski, Phys. Rev. \textbf{D87}, 094509 (2013),
  \texttt{1302.2249}

\bibitem{Mollgaard:2013qra}
A.~Mollgaard, K.~Splittorff, Phys. Rev. \textbf{D88}, 116007 (2013),
  \texttt{1309.4335}

\bibitem{Mollgaard:2014mga}
A.~Mollgaard, K.~Splittorff, Phys. Rev. \textbf{D91}, 036007 (2015),
  \texttt{1412.2729}

\bibitem{Bloch:2017sex}
J.~Bloch, J.~Glesaaen, J.J.M. Verbaarschot, S.~Zafeiropoulos, JHEP \textbf{03},
  015 (2018), \texttt{1712.07514}

\bibitem{Sexty:2013ica}
D.~Sexty, Phys. Lett. \textbf{B729}, 108 (2014), \texttt{1307.7748}

\bibitem{Aarts:2014bwa}
G.~Aarts, E.~Seiler, D.~Sexty, I.O. Stamatescu, Phys. Rev. \textbf{D90}, 114505
  (2014), \texttt{1408.3770}

\bibitem{Aarts:2016qrv}
G.~Aarts, F.~Attanasio, B.~Jäger, D.~Sexty, JHEP \textbf{09}, 087 (2016),
  \texttt{1606.05561}

\bibitem{Nishimura:2019qal}
J.~Nishimura, A.~Tsuchiya, JHEP \textbf{06}, 077 (2019), \texttt{1904.05919}

\bibitem{Hayata:2014kra}
T.~Hayata, A.~Yamamoto, Phys. Rev. \textbf{A92}, 043628 (2015),
  \texttt{1411.5195}

\bibitem{Berger:2018xwy}
C.~Berger, J.~Drut, PoS \textbf{LATTICE2018}, 244 (2018)

\bibitem{Attanasio:2019plf}
F.~Attanasio, J.E. Drut, Phys.\ Rev.\ A \textbf{101}, 033617 (2020),
  \texttt{1908.02715}

\bibitem{Loheac:2017yar}
A.C. Loheac, J.E. Drut, Phys. Rev. \textbf{D95}, 094502 (2017),
  \texttt{1702.04666}

\bibitem{Loheac:2018yjh}
A.C. Loheac, J.~Braun, J.E. Drut, Phys. Rev. \textbf{D98}, 054507 (2018),
  \texttt{1804.10257}

\bibitem{Rammelmuller:2018hnk}
L.~Rammelm{\"u}ller, A.C. Loheac, J.E. Drut, J.~Braun, Phys. Rev. Lett.
  \textbf{121}, 173001 (2018), \texttt{1807.04664}

\bibitem{Rammelmuller:2017vqn}
L.~Rammelm{\"u}ller, W.J. Porter, J.E. Drut, J.~Braun, Phys. Rev. \textbf{D96},
  094506 (2017), \texttt{1708.03149}

\bibitem{Rammelmuller:2020vwc}
L.~Rammelmüller, J.E. Drut, J.~Braun (2020), \texttt{2003.06853}

\bibitem{Shill:2018tan}
C.~Shill, J.~Drut, Phys.\ Rev.\ A \textbf{98}, 053615 (2018),
  \texttt{1808.07836}

\bibitem{Basu:2018dtm}
P.~Basu, K.~Jaswin, A.~Joseph, Phys. Rev. D \textbf{98}, 034501 (2018),
  \texttt{1802.10381}

\bibitem{Joseph:2019sof}
A.~Joseph, A.~Kumar, Phys. Rev. D \textbf{100}, 074507 (2019),
  \texttt{1908.04153}

\bibitem{Aarts:2013uza}
G.~Aarts, P.~Giudice, E.~Seiler, Annals Phys. \textbf{337}, 238 (2013),
  \texttt{1306.3075}

\bibitem{Aarts:2012ft}
G.~Aarts, F.A. James, J.M. Pawlowski, E.~Seiler, D.~Sexty, I.O. Stamatescu,
  JHEP \textbf{03}, 073 (2013), \texttt{1212.5231}

\bibitem{Cai:2019vmt}
Z.~Cai, Y.~Di, X.~Dong, Commun.\ Comput.\ Phys. \textbf{27}, 1344 (2020),
  \texttt{1905.11683}

\bibitem{Nagata:2015uga}
K.~Nagata, J.~Nishimura, S.~Shimasaki, PTEP \textbf{2016}, 013B01 (2016),
  \texttt{1508.02377}

\bibitem{Nishimura:2015pba}
J.~Nishimura, S.~Shimasaki, Phys. Rev. \textbf{D92}, 011501 (2015),
  \texttt{1504.08359}

\bibitem{Aarts:2017vrv}
G.~Aarts, E.~Seiler, D.~Sexty, I.O. Stamatescu, JHEP \textbf{05}, 044 (2017),
  \texttt{1701.02322}

\bibitem{Splittorff:2014zca}
K.~Splittorff, Phys. Rev. \textbf{D91}, 034507 (2015), \texttt{1412.0502}

\bibitem{Greensite:2014cxa}
J.~Greensite, Phys. Rev. \textbf{D90}, 114507 (2014), \texttt{1406.4558}

\bibitem{Seiler:2020mkh}
E.~Seiler (2020), \texttt{2006.04714}

\bibitem{Nagata:2018net}
K.~Nagata, J.~Nishimura, S.~Shimasaki, JHEP \textbf{05}, 004 (2018),
  \texttt{1802.01876}

\bibitem{Tsutsui:2019suq}
S.~Tsutsui, Y.~Ito, H.~Matsufuru, J.~Nishimura, S.~Shimasaki, A.~Tsuchiya
  (2019), \texttt{1912.00361}

\bibitem{Scherzer:2018hid}
M.~Scherzer, E.~Seiler, D.~Sexty, I.O. Stamatescu, Phys. Rev. \textbf{D99},
  014512 (2019), \texttt{1808.05187}

\bibitem{cai2020validity}
Z.~Cai, X.~Dong, Y.~Kuang (2020), \texttt{2007.10198}

\bibitem{Scherzer:2019lrh}
M.~Scherzer, E.~Seiler, D.~Sexty, I.O. Stamatescu, Phys. Rev. \textbf{D101},
  014501 (2020), \texttt{1910.09427}

\bibitem{Banerjee:2010kc}
D.~Banerjee, S.~Chandrasekharan, Phys. Rev. \textbf{D81}, 125007 (2010),
  \texttt{1001.3648}

\bibitem{Hirasawa:2020bnl}
M.~Hirasawa, A.~Matsumoto, J.~Nishimura, A.~Yosprakob (2020),
  \texttt{2004.13982}

\bibitem{UBHD-68404032}
M.~Scherzer, \emph{PhD thesis} (Heidelberg, 2019)

\bibitem{Attanasio:2018rtq}
F.~Attanasio, B.~Jäger, Eur. Phys. J. \textbf{C79}, 16 (2019),
  \texttt{1808.04400}

\bibitem{Nagata:2018mkb}
K.~Nagata, J.~Nishimura, S.~Shimasaki, Phys.\ Rev.\ D \textbf{98}, 114513
  (2018), \texttt{1805.03964}

\bibitem{Sexty:2019vqx}
D.~Sexty, Phys. Rev. \textbf{D100}, 074503 (2019), \texttt{1907.08712}

\bibitem{Ito:2018jpo}
Y.~Ito, H.~Matsufuru, J.~Nishimura, S.~Shimasaki, A.~Tsuchiya, S.~Tsutsui, PoS
  \textbf{LATTICE2018}, 146 (2018), \texttt{1811.12688}

\bibitem{Kogut:2019qmi}
J.B. Kogut, D.K. Sinclair, Phys. Rev. \textbf{D100}, 054512 (2019),
  \texttt{1903.02622}

\bibitem{Ito:2020mys}
Y.~Ito, H.~Matsufuru, Y.~Namekawa, J.~Nishimura, S.~Shimasaki, A.~Tsuchiya,
  S.~Tsutsui (2020), \texttt{2007.08778}

\bibitem{Scherzer:2020kiu}
M.~Scherzer, D.~Sexty, I.O. Stamatescu (2020), \texttt{2004.05372}

\bibitem{Morningstar:2003gk}
C.~Morningstar, M.J. Peardon, Phys. Rev. \textbf{D69}, 054501 (2004),
  \texttt{hep-lat/0311018}

\bibitem{Kogut:2007mz}
J.~Kogut, D.~Sinclair, Phys. Rev. D \textbf{77}, 114503 (2008),
  \texttt{0712.2625}

\bibitem{Bloch:2017jzi}
J.~Bloch, O.~Schenk, EPJ Web Conf. \textbf{175}, 07003 (2018),
  \texttt{1707.08874}

\end{thebibliography}

\end{document}